# Galactic Component Mapping of Galaxy UGC 2885 by Machine Learning Classification

Robin J. Kwik[1,3*], Jinfei Wang[1,3], Pauline Barmby[2,3], Benne W. Holwerda[4]

[1]Department of Geography and Environment, University of Western Ontario, London, ON N6A 5C2, Canada
[2]Department of Physics and Astronomy, University of Western Ontario, London, ON N6A 3K7, Canada
[3]The Institute for Earth and Space Exploration, University of Western Ontario, London, ON N6A 3K7, Canada
[4]Department of Physics and Astronomy, University of Louisville, Louisville, KY 40292

\* Corresponding author.
  *E-mail address*: rkwik@uwo.ca (R. Kwik)

## Abstract

Automating classification of galaxy components is important for understanding the formation and evolution of galaxies. Traditionally, only the larger galaxy structures such as the spiral arms, bulge, and disc are classified. Here we use machine learning (ML) pixel-by-pixel classification to automatically classify all galaxy components within digital imagery of massive spiral galaxy UGC 2885. Galaxy components include young stellar population, old stellar population, dust lanes, galaxy center, outer disc, and celestial background. We test three ML models: maximum likelihood classifier (MLC), random forest (RF), and support vector machine (SVM). We use high-resolution Hubble Space Telescope (HST) digital imagery along with textural features derived from HST imagery, band ratios derived from HST imagery, and distance layers. Textural features are typically used in remote sensing studies and are useful for identifying patterns within digital imagery. We run ML classification models with different combinations of HST digital imagery, textural features, band ratios, and distance layers to determine the most useful information for galaxy component classification. Textural features and distance layers are most useful for galaxy component identification, with the SVM and RF models performing the best. The MLC model performs worse overall but has comparable performance to SVM and RF in some circumstances. Overall, the models are best at classifying the most spectrally unique galaxy components including the galaxy center, outer disc, and celestial background. The most confusion occurs between the young stellar population, old stellar population, and dust lanes. We suggest further experimentation with textural features for astronomical research on small-scale galactic structures.

## 1. Introduction

Galaxies contain stars, dust, and gas that make up larger bulge, bar, disc, and spiral arm structures. Nearby galaxies – galaxies in close enough proximity that their components can be resolved in high resolution imagery – are particularly useful for study of the spatial and temporal relationships. By studying galaxies, we can learn about the structure and evolution of our own Milky Way galaxy. Studying the evolution of galaxies is an important step to understanding how all matter formed. Further, by focusing on nearby galaxies we can apply our gained knowledge to more distant galaxies that we cannot observe in full detail (Bianchi et al., 2014; Kalirai, 2018). It is also known that deconstruction of spiral galaxies into their components plays a key role in understanding the nature of galactic evolution and the structural properties of components such as stars, dust, and gas (Lingard et al., 2020). Therefore, studying the spatial distribution of galactic components will broaden the understanding of the photometric properties of galaxies. By automating the process of galactic component analysis, the same methodologies can be applied to high-resolution imagery of similar nearby galaxies such as the Legacy ExtraGalactic Ultraviolet Survey (LEGUS) and the Physics at High Angular Resolution in Nearby GalaxieS with the Hubble Space (PHANGS-HST) survey (Calzetti et al., 2015; Lee et al., 2021).

Bulge-disc decomposition of galaxies – the photometric or spectroscopic separation of the bulge and disc regions of a spiral galaxy within digital imagery – is a well researched area of study (a recent example: Pak et al., 2021). However, to our knowledge, identification and mapping of all galactic components within a galaxy's bulge and disc has not been done. Although an expert can perform visual inspection and identification of galaxy components, the automation of this process can shorten the time required to analyze digital imagery of galaxies. By classifying each component within digital imagery of spiral galaxies, the fine details we observe and identify are remnants of star formation history of the galaxy and can provide clues to the evolution and formation of matter within the galaxy (Peng et al., 2002). The use of human visual interpretation is not a new idea, but no study has used this method for pixel-by-pixel component identification of high resolution imagery. Citizen science based projects such as Galaxy Zoo (Lingard et al., 2020; Lintott et al., 2008) work to expand upon the traditional bulge-disc classification by having participants identify components within digital imagery of galaxies. However, these studies focus on identifying sub-structures within the bulge and disc such as spiral arms and bars. The success of human visual interpretation to identify galaxy structures demonstrates the usefulness of visual observation of digital imagery. By surpassing the traditional bulge-disc decomposition and digging deeper into galactic structure, we can aid in the quantitative understanding of galactic component distribution and evolution (Lingard et al., 2020).

In recent years, there has been growing interest in machine learning (ML) for digital image analysis (Baron, 2019; Thanh Noi & Kappas, 2017). By supervising the identification of pixels that represent different classes, the machine learns the photometric characteristics of classes and can then automatically classify individual pixels in digital imagery based on the algorithm's acquired knowledge. Several recent studies have used pixel-based ML to classify galaxies. Hausen & Robertson (2020) use ML methods to classify morphologies or types of galaxies within a Hubble image by use of a pixel-based method. Bialopetravičius & Narbutis (2020) use ML to identify star clusters within a nearby nearly face-on galaxy. Both studies emphasize the need for automation of high resolution galaxy classification methods. However, their focus on morphology and star clusters neglects other significant galaxy components such as dust lanes and stellar populations that are not members of clusters. This knowledge gap is significant as pixel based classification of galactic populations is necessary for mapping complex distributions of galactic material and for better understanding the complex relationships within galaxies. Further, the upcoming Euclid (Laureijs et al., 2011) and Roman Space Telescope (Spergel et al., 2015) will make available more high-resolution Hubble-like data for nearby galaxies, making pixel-based mapping more feasible.

Texture analysis is a commonly used image processing technique in earth-based remote sensing. Several studies have explored the usefulness of texture for morphological analysis of galaxies (Au, 2006; Ntwaetsile & Geach, 2021; Pedersen et al., 2013; Schutter & Shamir, 2015; Shamir, 2009; Shamir, 2021). Ntwaetsile & Geach (2021) find that texture analysis is particularly useful for radio galaxy morphology analysis and recommend that it be applied to large imaging surveys. Similarly, Shamir (2021) notes the usefulness of texture analysis for identifying outlier galaxies in optical imagery of galaxies. Both examples demonstrate the diverse applications of texture for astronomical imagery. However, to our knowledge texture has yet to be tested for identification of galactic components within digital imagery. Because texture is useful for summarizing imagery, we expect texture to be particularly useful for identifying the differences between the fine details of galaxy components within high resolution digital imagery.

A geographic information system (GIS) method not commonly used in classification of celestial phenomena is distance as calculated for the pixel contents of a digital imagery. Because components of galaxies are arranged based on spatial and temporal patterns, distance measures commonly used in GIS for Earth-based phenomena are compatible with these galaxies. In one particular instance of use of distance for astronomical research, Bialopetravičius & Narbutis (2020) make use of distance from galaxy center and spiral arms to observe the relationship between distance from spiral arms and galaxy center to the age of stars within galaxy M83. However, the distance measures are implemented for post-classification analysis only. To our knowledge no study has incorporated per pixel distance measures within the galaxy plane to classify all galaxy components within digital imagery.

By addressing the below research objectives, we hope to better understand dynamics of nearly edge-on nearby galaxies and the most efficient method of identifying galactic components within digital imagery through machine learning classification:

1. Evaluate how machine learning can be used to classify galactic components,
2. Determine which input parameters are most useful,
3. Determine which machine learning method is most useful.

## 2. Methods
### 2.1. Study Area

Galaxy UGC 2885 or 'Rubin's Galaxy' is an unusually large and late type (Sc) spiral galaxy sitting 79.1 Mpc or approximately 258,000,000 light years away (Hunter et al., 2013; Figure 1). Rubin's Galaxy is at a suitable distance for observation of galactic components; at these distances, the stellar field is so dense that most pixels contain the integrated light from multiple stars. The massive size of UGC 2885 –approximately 44.4 kpc or 145,000 light years in diameter (Hunter et al., 2013) – makes it an interesting case study for mapping of the populations within it. By studying the distribution of galactic components within UGC 2885, we can better understand the nature of spatial and temporal patterns within the galaxy, as well as how massive galaxies differ from galaxies with more common properties. UGC 2885 is also inclined 74° (Hunter et al., 2013), meaning that it is nearly edge-on, where a completely edge-on galaxy is 90°. The inclination of a galaxy is defined relative to the point of view of an observer.

    At the center of UGC 2885, a supermassive black hole has been identified (Holwerda et al., 2021). The massive size of UGC 2885, the presence of a supermassive black hole, and the lack of star formation within the galaxy make UGC 2885 defy easy morphological classification. As Holwerda et al. (2021) note, classifying the small structures within the galaxy can contribute to a better understanding of UGC 2885 and similarly unusual galaxies. Ultimately, UGC 2885 is an optimal study area for galactic component mapping due to the large population of components within the galaxy. This particular galaxy also exhibits spatial and temporal gradients of stars. For instance, the galactic center is redder due to the large number of old stars within it. On the contrary, the spiral arms have a larger concentration of young stellar populations making them appear bluer.

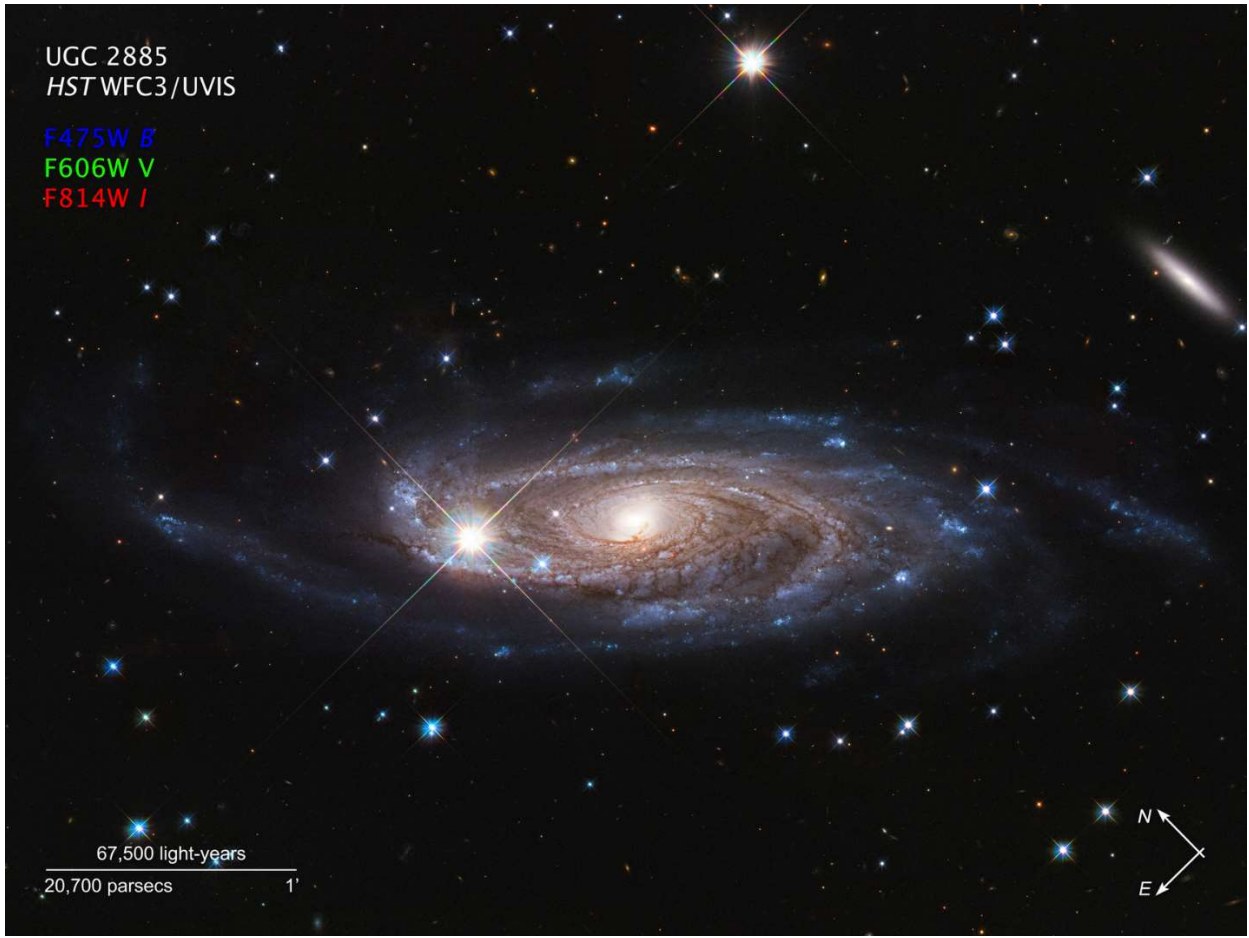

**Figure 1.** HST colour composite map of UGC 2885 where the F475W band is displayed in blue, the F606W band is displayed in green, and the F814W band is displayed in red. This image is in celestial orientation; therefore, east is towards the left rather than to the right of the north direction. Parsecs refers to a measure of distance equalling approximately 3.26 light years. Image credit: NASA, ESA, and B.W. Holwerda (University of Louisville).

## 2.2. Data

We use the Hubble Space Telescope (HST) multispectral digital imagery – the highest resolution imagery currently available in visible and near infrared wavelengths – for UGC 2885 in three wavelength bands. Holwerda et al. (2020) generously provide mosaics for all three wavelength bands. Observations in three wavelength bands – band F475W, F606W, and F814W – are available as part of HST program 15107, *The Cluster Population of UGC 2885* (Holwerda, 2017)[1]. Among the three bands, band F475W shows blue-green (B) emission from the galaxy (Fukugita et al., 1996), band F606W shows visual (V) emission from the galaxy, and band F814W shows near-infrared (I) emission from the galaxy (Dressel, 2021). These bands are significant for mapping stellar populations and other galactic components within nearby galaxies (Holwerda et al., 2020; Kiar et al., 2017). More specifically, the B band is useful for observing younger and hotter stars while the V and I bands are useful for identifying the cooler and redder stars (Kiar et al., 2017). The BVI bands can also be used to observe dust lanes throughout the galaxy; in particular, because the dust lanes are redder in wavelength, the dust lanes emit stronger in the V and I bands. Therefore, the HST BVI bands are useful for observation of the major galaxy components within the UGC 2885.

The HST mosaics (Holwerda et al., 2020) are in Flexible Image Transport System (FITS) format. We use FITS Liberator 3 software (ESA/ESO/NASA, 2021) to export the images to Tag Image File Format (TIFF) using a logarithmic stretch. The logarithmic stretch is for visualization purposes only and does not change the original values of the HST imagery.

Due to the 74° inclination of UGC 2885, we also deproject the digital imagery to approximate a face on galaxy to calculate Euclidean distance as described in section 2.3.4. (Figure 2). We use deprojection only for distance layer generation. To deproject the digital imagery, we calculate a trigonometric stretch ratio of $1/\cos(74°)$ and find a value of 3.628. Therefore, we stretch the digital imagery vertically by 363% using a raster graphics editor; see also Davis et al. (2012) for an excellent description of deprojection of similarly inclined galaxies. Due to the presence of foreground stars in the HST imagery, we also mask the 20 brightest stars within a three arcminute radius of UGC 2885 from Gaia Early Data Release 3 (EDR3) data (Gaia Collaboration et al., 2016; Gaia Collaboration et al., 2021).

---

[1] The band number (i.e. 475) represents the central wavelength of the sensitivity range of the band.

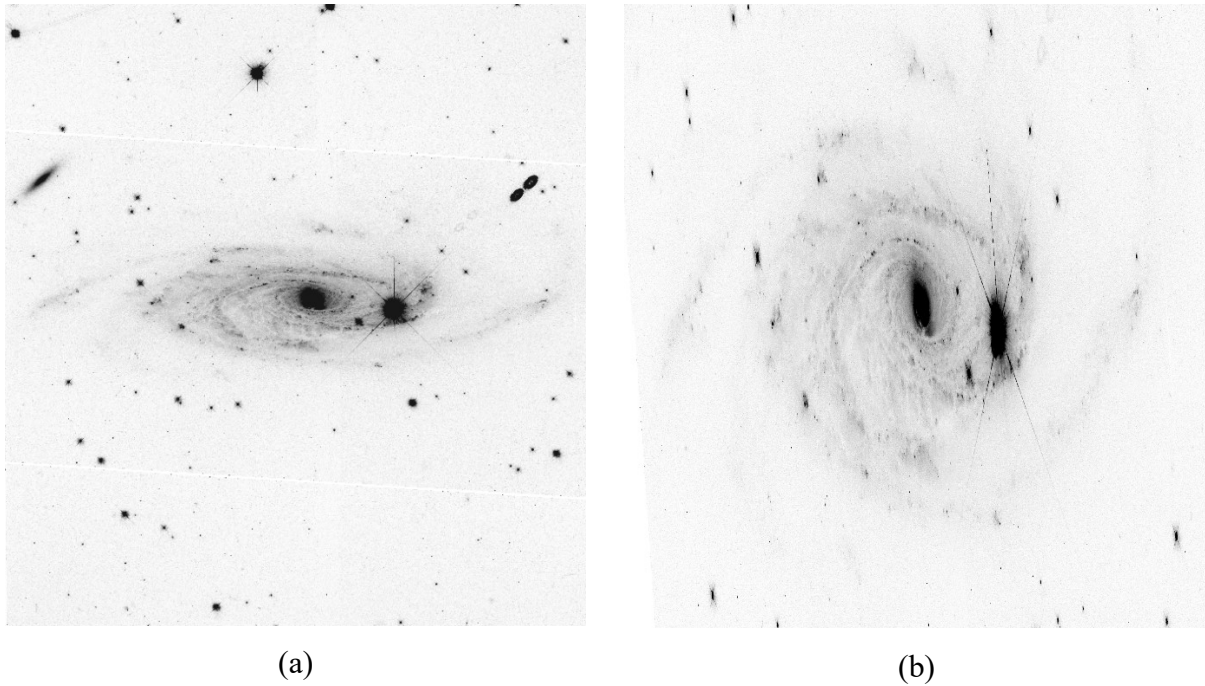

(a)            (b)

**Figure 2.** UGC 2885 blue-green band imagery (F475W): a) original HST imagery showing the projected galaxy with an inclination of 74°; b) deprojected image of UGC 2885 – stretched vertically by 363%.

## 2.3. HST Image Processing
### 2.3.1. Coordinate Transformation

For spatial analysis in a GIS, imagery must be transformed to an Earth-based coordinate system. We retrieve coordinates for features within the image using SAOImage DS9 (Version 8.0.1.) open source software (Joye & Mandel, 2003). HST images of UGC 2885 are transformed using the Helmert transformation (Farhadian & Clarke, 2020) available within the "Georeferencer" tool in an open source software QGIS Version 3.16.3. (QGIS 3.16., 2021). The Helmert transformation shifts and rotates the image through affine methods that preserve the collinearity and ratio of distance of the features in an image (Zhili Song et al., 2014). Testing other transformations in QGIS resulted in highly distorted imagery, leading us to use the Helmert transformation.

     Figure 3 shows the HST image processing methodology following coordinate transformation. With the transformed HST imagery, we create an image stack in GIS software.

From the HST raster composite, we produce textural features, band ratios, and distance layers. We further describe the methodology (Figure 3) in the following sections.

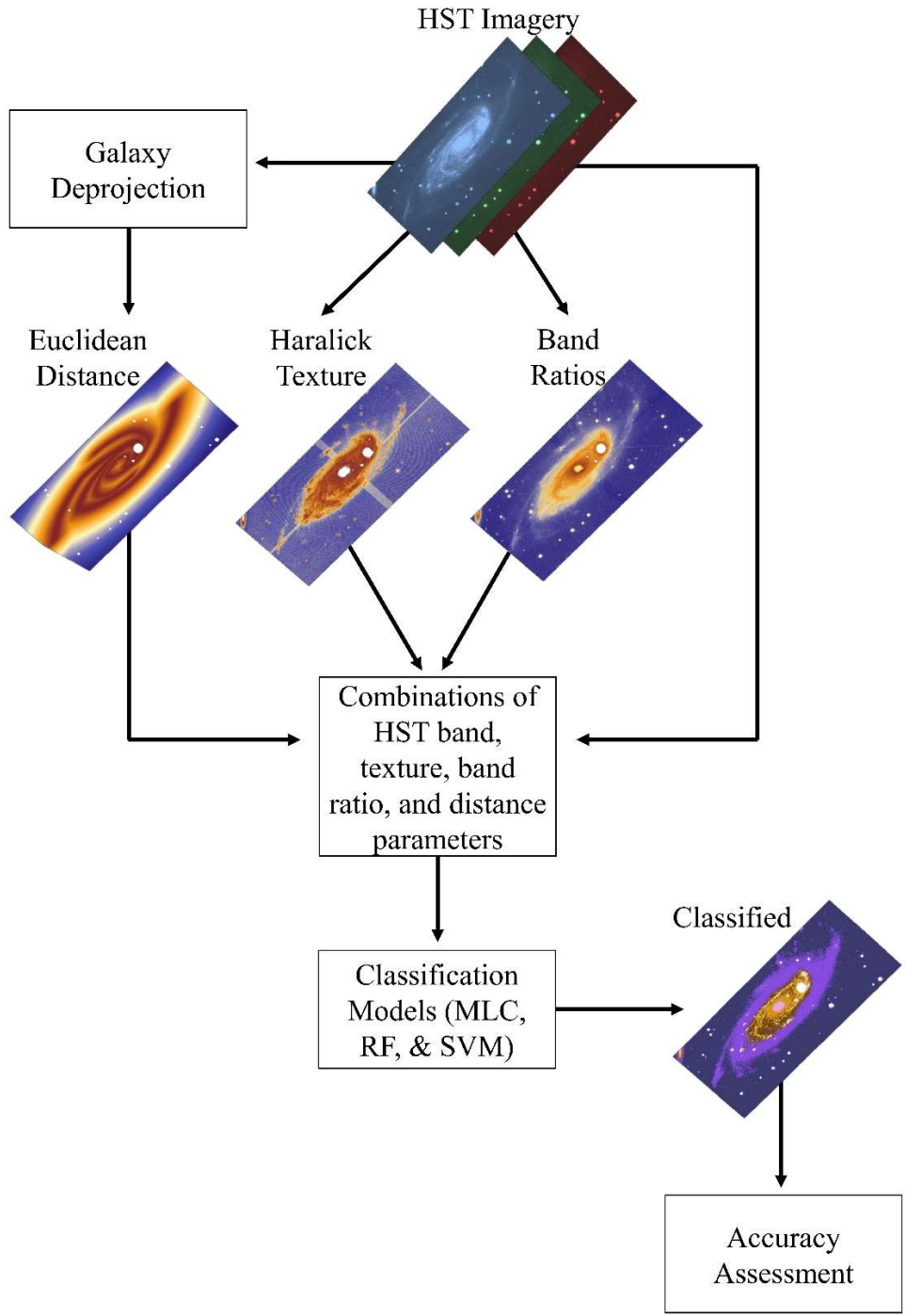

**Figure 3.** Flowchart showing methods used in this study.

### 2.3.2. Band Ratio

Band ratios are effective as they are useful at describing how stars emit light and how absorption affects that light. For instance, bluer stars are more affected by dust lane absorption so will have different ratio values than a red star affected by the same dust. We use several broadband ratios commonly used in astronomy to identify particular aspects of UGC 2885. Kiar et al. (2017) provide a good overview of HST band ratios. We specify similar band ratios for UGC 2885 in Table 1 and expect the ratios to emphasize different phenomena such as star clusters within the galaxy. We calculate band ratios by division of wavelength bands in a GIS software. Typically, the shorter wavelength band by is divided by the longer wavelength band when calculating simple ratios (i.e. B/V). Band ratios used in classification are described below.

**Table 1.** Band ratios used for classification and their respective equations.

| Flux Band Ratio | Fraction Band Ratio | Spectral Slope Band Ratio |
| --- | --- | --- |
| B-V | B/(B+V+I) | (B/V)/(B+V+I) |
| B-I | V/(B+V+I) | (B/I)/(B+V+I) |
| V-I | I/(B+V+I) | (V/I)/(B+V+I) |

Before creating ratios, we calibrate the HST bands F475W, F606W, and F814W with their respective Flexible Image Transport System (FITS) calibration factors found in the Imager Header tab within FITS Liberator 3 software (ESA/ESO/NASA, 2021; Table 2). The process of calibration involves multiplying each pixel in a raster image by a certain value in order to convert the pixels into meaningful units; we convert the original 32-bit pixel range to Jansky units which are equal to $10^{-26}$ Watts metre$^{-2}$ Hertz$^{-1}$. For instance, we use Eq. (1) to calibrate the F475W blue-green band:

*Calibrated Raster* = "F475W raster" * 1.8782514E-07    (1)

We repeat the above calculation for both the F606W and F814W bands. After calibration, we compute the flux, fraction, and spectral slope band ratios.

**Table 2.** Calibration factors for the BVI imagery as found in FITS Liberator 3 Image Headers tab under "PHOTFNU".

| Band | Calibration Factor |
|---|---|
| F475W (Blue-green) | 1.8782514E-07 / Inverse sensitivity, Jy*sec/e- |
| F606W (Visual) | 1.32242795E-07 / Inverse sensitivity, Jy*sec/e- |
| F814W (Near-Infrared) | 3.2380001E-07 / Inverse sensitivity, Jy*sec/e- |

Flux band ratios involve simple band division. Three combinations of flux band ratios are possible with HST data available for UGC 2885: B-V, B-I, and V-I. For the B-V ratio, we perform division of blue-green and visual bands (B/V) in a GIS software where the highest pixel values are those having a higher value in the blue-green band and a lower value in the visual band. The same concept applies to the B-I and V-I ratios.

To calculate fraction ratios, we divide a band by the flux sum of all three bands. For instance, B / (B+V+I) calculates the fraction of blue-green-to-total light emitted from UGC 2885. The fraction band ratios compare the band flux of a single band to the total brightness, therefore they identify the most prominent emission wavelengths for a given pixel. For instance, the B-fraction band ratio (B/(B+V+I)) will identify the bluest sources within the galaxy, these being the young stars. B-fraction band ratio also identifies dust lanes. Dust is a better absorber of blue light than red light, so the dust appears darkest in the B band. The V band shows a wider range of visual light, therefore the V-fraction band ratio (V/(B+V+I)) shows many galaxy components. In particular, it does a good job of emphasizing the structure of the galaxy in both the inner and outer disc that contain old and young stars. We can also make out clear dust lanes in the V-fraction band ratio image. Near-infrared (I) will have a larger emission of light for the old stars, dust lanes, and galaxy center, so the I-fraction band ratio (I/(B+V+I)) emphasizes the inner disc of the galaxy where the old stars are accumulated.

The final band ratio, spectral slope, calculates the spectral slope of two bands over the flux product of the three bands. For example, (B/V) / (B+V+I), calculates the blue-green/visual slope. The spectral slope ratio accounts for any correlation between colour, as measured by band ratios, and overall brightness, as measured by the band sum, making them akin to colour-magnitude diagrams used in astronomy. These quantities might be expected to correlate because dust within a galaxy will make the emergent light both fainter and redder. Colour and brightness are direct indicators of stellar age: young stars are brighter and bluer while older stars are dimmer and redder. The BV spectral slope ratio ((B/V)/(B+V+I)) shows the most detail within the inner disc whereas in the BI spectral slope ratio ((B/I)/(B+V+I)) image it is more difficult to distinguish the galaxy center. The VI spectral slope ratio ((V/I)/(B+V+I)) image places most emphasis on the galaxy center region.

### 2.3.3. Texture Features

Texture features imitate visual patterns we see in objects, area, and phenomena. We calculate Haralick Grey Level Co-occurrence Matrix (GLCM) textures (Haralick et al., 1973) for HST imagery of UGC 2885. The GLCM is produced from all pixel grey level values within a moving window of specified size and considers the grey levels of two pixels at a time, the reference pixel and the neighbouring pixel. For instance, a 5x5 window will produce a GLCM from 25 pixels. The number of grey levels chosen by the user determines the size of the GLCM, and the pixel values of the original imagery are scaled down to the chosen number of grey levels. On the position of the GLCM where the grey levels of the reference and neighbour pixel meet, 1 is added to that position. After the GLCM is produced, second order statistics are calculated based off of the contents of the GLCM (Haralick et al., 1973).

In galactic imagery, dust lanes often exhibit a rougher texture while the bright galaxy center and star clusters have a smooth appearance. We aim to determine if texture features can identify these differences and increase accuracy of classification. GLCM textures have been successfully tested for many remote sensing research applications of Earth-based phenomena (Ghasemian & Akhoondzadeh, 2018; Hall-Beyer, 2017; Wei et al., 2021; Zhang et al., 2014), making texture a promising prospect for astronomical research. Textures are excellent at rapidly summarizing the contents of an image, so are useful for processing the abundance of astronomical data for machine learning classification (Ntwaetsile & Geach, 2021).

We use eight Haralick textures (Haralick et al., 1973) available in a commercial remote sensing software; textures include angular second moment, contrast, correlation, dissimilarity, entropy, homogeneity, mean, and standard deviation. We specify 64 grey levels and a sliding window size of 5x5. Any grey level can be chosen, but we specify 64 as we find it suitable for the HST imagery. Further, we choose a window size of 5x5 pixels based on the measured pixel width of a typical star cluster within the HST BVI imagery. We describe textures below and show their appearance when calculated for HST band F606W in Figure 4.

Entropy texture calculates unevenness of the image grey levels respectively (Wei et al., 2021) and is represented by the following equation:

$$Entropy = -\sum_{i=1}^{L}\sum_{j=1}^{L} \hat{P}(i,j) \log[\hat{P}(i,j)] \qquad (2)$$

where P(i,j) is the (i,j)th entry in a normalized GLCM. Entropy texture is high when an image has a large range of grey levels, therefore having unevenness.

Angular second moment calculates textures based on uniformity of the imagery. Eq. (3) describes how angular second moment is calculated:

$$Angular\ second\ moment = \sum_{i=1}^{L}\sum_{j=1}^{L}(\hat{P}(i,j))^2 \qquad (3)$$

Images with a larger number of grey levels have smaller uniformity and therefore smaller values of angular second moment texture.

Homogeneity looks at the evenness or homogeneous nature of the spectral characteristics throughout an image and is calculated by the following Eq. (4):

$$Homogeneity = \sum_{i=1}^{L}\sum_{j=1}^{L} \frac{\hat{P}(i,j)}{1+(i-j)^2} \qquad (4)$$

Mean calculates the average of the grey-levels in the GLCM local window and is defined by Eq. (5):

$$Mean = \sum_{i=1}^{L}\sum_{j=1}^{L} i * \hat{P}(i,j) \qquad (5)$$

Contrast and dissimilarity are very similar to each other in that they both measure the spectral variation within the local GLCM window. However, they are different in that contrast incorporates the square root of the difference between *i* and *j* co-occurrence matrix where dissimilarity uses the absolute difference between *i* and *j* co-occurrence matrix. Contrast is defined as:

$$Contrast = \sum_{n=0}^{L-1}(i-j)^2 \sum_{i=1}^{L}\sum_{j=1}^{L} \hat{P}(i,j) \qquad (6)$$

where high contrast indicates textures with sharp edges in an image. Dissimilarity is described by the following equation:

$$Dissimilarity = \sum_{i=1}^{L}\sum_{j=1}^{L}|i-j|\hat{P}(i,j) \qquad (7)$$

Standard deviation texture calculates the standard deviation or scattering of the local spectral information with respect to the mean. Areas where pixels have a small range of values have lower standard deviation while areas with high pixel ranges have higher standard deviation. Eq. (8) defines standard deviation:

$$Standard\ Deviation = \sum_{i=1}^{L}\sum_{j=1}^{L}(i-\mu)^2 \hat{P}(i,j) \qquad (8)$$

Correlation is a measure of linear dependency of the spectral variation on local pixels within the GLCM window. High values show areas where noise or sharp changes are present in the image. Correlation is described by Eq. (9):

$$Correlation = \sum_{i=1}^{L}\sum_{j=1}^{L}((i*j*\hat{P}(i,j) - \mu_x\mu_y)/\sigma_x\sigma_y) \qquad (9)$$

where $\mu_x$ and $\mu_y$ are the means of $P_x$ and $P_y$, and $\sigma_x$ and $\sigma_y$ are the standard deviations of $P_x$ and $P_y$.

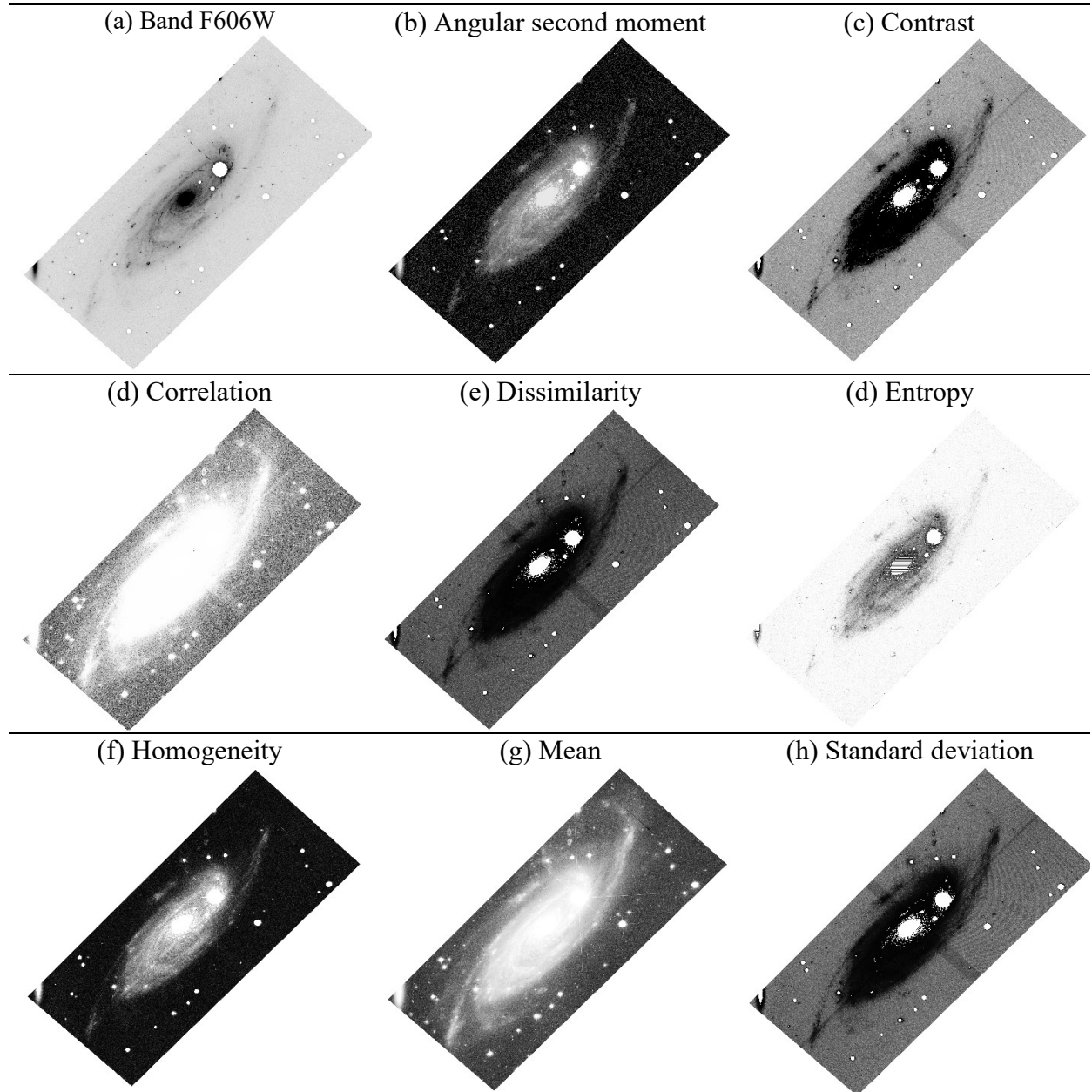

**Figure 4.** Band F606W Haralick textures with 20 brightest foreground stars masked.

### 2.3.4. Distance from Spiral Arms and Galaxy Center

Distance information is useful in spiral galaxies that exhibit age gradients. Galaxies that are dominated by spiral density waves (Lin & Shu, 1964) are most compatible with measurements of distance. Density waves are thought to be present in UGC2885 (Canzian et al., 1993), meaning distance measures are expected to be useful for classification of galaxy components. Young stars typically form in the dense spiral arms and disperse as they age forming the age gradient. The use of distance information can help to better understand the existence of age gradients within UGC2885.

Tracing of spiral arms has been used in galaxies with prominent dust lanes along their spiral arms (Shabani et al., 2018). However, galaxy UGC 2885 has some ambiguity in the spiral arm structure due to a lack of observable dust lanes in the optical HST imagery and the galaxy's 74° inclination. Therefore, we define the spiral arms by fitting logarithmic spirals, which are good approximations of the shape of spiral arms (Davis et al., 2012; Seigar & James, 1998). To ensure our tracing of spiral arms follows logarithmic structure, we perform a piece-wise fit of the spiral arms by manually overlaying logarithmic spirals onto the HST imagery and selecting segments. This is done in Desmos, a free online graphing program. When we identify a sufficient piece-wise fit, we plot points along the spiral arms, connecting the points using a parametric curve. We also draw a polygon over the galaxy center in the deprojected HST imagery.

The tool used to compute Euclidean distance converts the vector spiral arm line features to raster by generating pixels along the spiral arm lines. Distance is therefore calculated for each cell in a specified extent to the closest raster pixel in the spiral arm line. Following distance calculation, the distance raster is clipped to match the extent of the HST imagery and foreground stars are masked. The inter-arm regions, those between the spiral arms, will have distance calculated from the nearest spiral arm by way of the shortest distance. The galaxy center feature is also converted to raster pixels, and Euclidean distance is calculated for each pixel to the nearest edge pixel in the galaxy center raster by way of the shortest distance. We end up with two distance layers in decimal degrees and multiply both rasters by 3600 to convert to distance in arcseconds. This step brings the raster values closer to the 32-bit range of values of the HST bands that we will input into classification along with the distance bands. We then reproject the distance layers to the original extent of the HST imagery. The spiral arm and galaxy center features as well as their respective distance layers are shown in Figure 5.

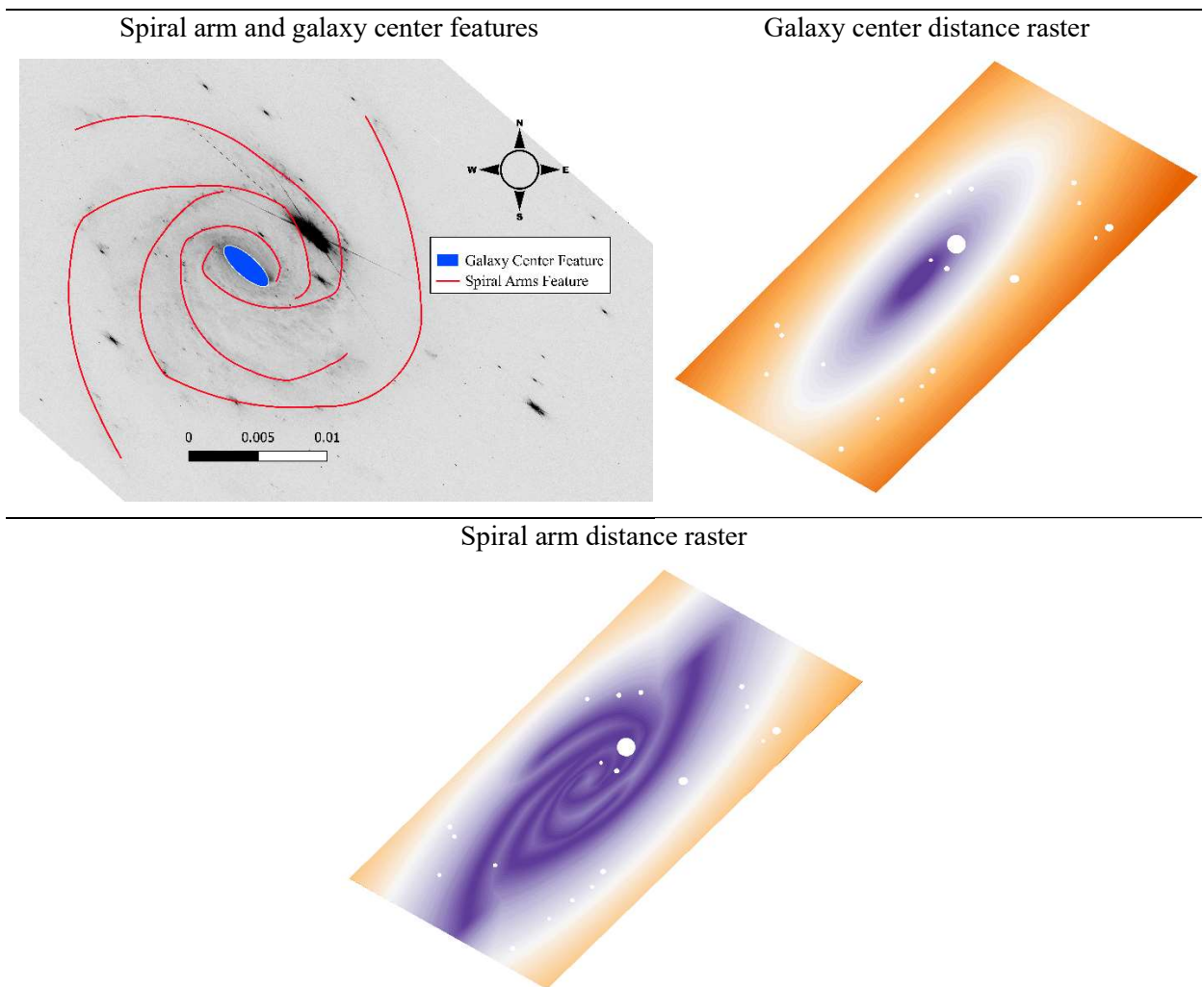

**Figure 5.** Spiral arms and galaxy center shown in red and blue respectively. The background image shows the deprojected UGC 2885 F475W blue-green band (Holwerda et al., 2020). Both distance rasters are reprojected to the galaxy's 74° inclination.

## 2.4. Classification Schemes

Although UGC 2885 is considered a nearby galaxy, it is not near enough for observation of individual stars. However, we can observe groups of stars called star clusters meaning that there is variation within galactic components we observe in the digital imagery. Because of UGC 2885's vast distance, we also have no access to 'ground truth' like we do for Earth-based phenomena. Therefore, we use our expert classification to train the models based off our theoretical understanding of galaxy components in the HST imagery. Along with HST imagery, we use the distance layers (galaxy center and spiral arm) and the band ratios to create training sites. The distance and band ratios act as complementary information to confirm the visual identification, and help to reduce subjectivity of training site creation.

To improve reproducibility we provide a guide to classification schemes used in this research. Figure 6 shows an example of how we defined the classification schemes within a GIS software. We decide on six classes based on their spectral values and visual appearance within the digital imagery: young stellar population (C1), old stellar population (C2), dust lanes (C3), galaxy center (C4), outer disc (C5), and celestial background (C6). Although the celestial background is not a part of the galaxy, we include it to avoid confusion with similar pixels within the galaxy.

We create training site polygons over areas of the digital imagery representing the six classes. Figure 7 visually demonstrates class separability between HST visual band F606W (Layer 2) and infrared band F814W (Layer 3) using a scatterplot of pixel values from each class. There is some class confusion due to the variation within galactic components in the HST imagery; the most confusion occurs between the old stellar populations and dust lanes that are similar in appearance in the HST imagery. Young and old stellar populations exhibit the widest range in pixel values. Because of saturation in the original HST imagery, the galaxy center and the young stellar population have saturated points concentrated in the top right of the graph.

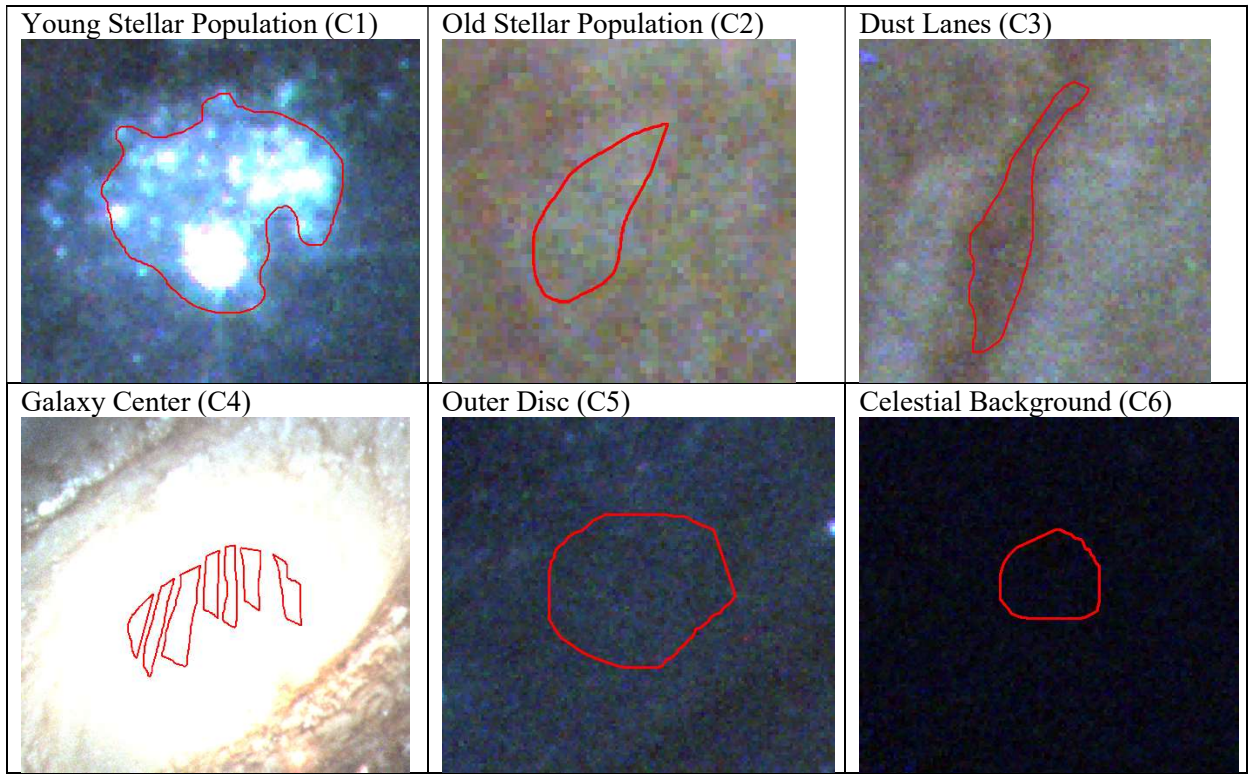

**Figure 6.** Examples of training site selection using the classification scheme with six classes. Background image is a RGB colour composite of HST band data of UGC 2885: F475W band is displayed in blue, the F606W band is displayed in green, and the F814W band is displayed in red.

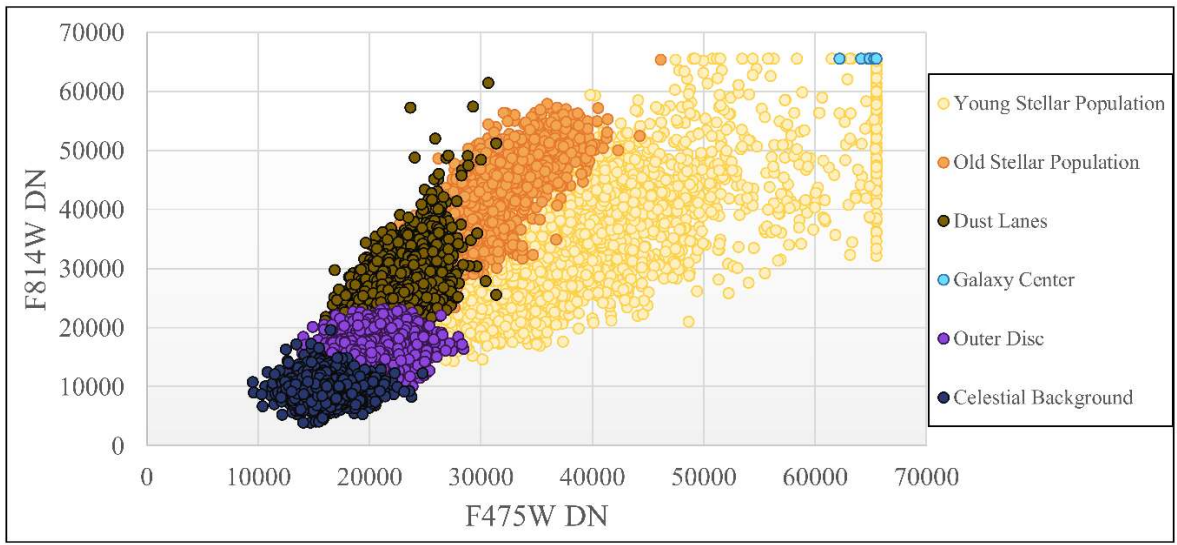

**Figure 7.** Scatterplot of classification scheme pixels and their respective digital number (DN) values within the visual and near-infrared HST bands.

## 2.5. Machine Learning Algorithms

To test the usefulness of machine learning for galactic component identification, we compare the performance of traditional Maximum Likelihood Classifier (MLC) model to the more powerful and increasingly popular Random Forest (RF) and Support Vector Machine (SVM) models. All three models are commonly used for image classification (Baron, 2019; Fluke & Jacobs, 2020; Lavallin & Downs, 2021; Maxwell et al., 2018).

### 2.5.1. Maximum Likelihood Classifier

The MLC model is capable of classifying pixels in an image into probability density functions based on their variance and covariance statistics (Foody et al., 1992; Norovsuren et al., 2019). In our case, MLC assigns each pixel in the imagery to one of the six classes specified. We performed MLC in a GIS. We can define MLC by the following equation (Richards & Jia, 1999):

$$D = \ln(a_c) - [0.5\ln(|COV_c|) - [0.5(X - M_c)T(COV_c - 1)(X - M_c)] \quad (11)$$

where $D$ is the likelihood, $c$ is the particular class in question, $COV_c$ is the covariance matrix for the class $c$ pixels, $X$ is a measurement vector for a specific pixel, and $Mc$ is a mean vector of a class ($c$).

### 2.5.2. Support Vector Machine

The final method of machine learning classification is SVM, a method that uses a hyperplane to define an optimal split between classes. An optimal split can be defined as one that separates the natural groupings in the samples while maintaining the maximal distance from support vectors, which are extreme samples within the data (Fluke & Jacobs, 2020). When training the SVM models, we use the value of 500 samples per class. We test SVM using several numbers of samples including 125, 250, 500, 750, and 1000. We find that the default value of 500 samples per class is sufficient, as there is no drastic change in accuracy when testing with the other numbers of samples. Shao and Lunetta (2012) find that SVM performs well with a low sample size of 20 pixels, although classification accuracy did increase when testing up to 800 pixels per class. SVM is not as sensitive to training sample sizes as RF (Thanh Noi & Kappas, 2017) so it makes sense the accuracy does not drastically change. We evaluate the accuracy of training sample sizes using overall accuracy statistic, which is a measure of the sum of the individual class accuracy (correctly classified pixels in each class) divided by the total number of pixels in the testing data.

### 2.5.3. Random Forest

The RF model is an ensemble algorithm that relies on a set of decision trees that each make a decision about the state of a sample. After processing a sample through its decision trees, the class or state of that sample is decided through a majority vote of the trees meaning that the class or state most commonly identified by the decision trees is assigned to the sample (Breiman, 1984; Breiman, 2001).

We performed RF in a GIS. From testing of models with 125, 250, 500, 750, and 1000 trees, we find that there is no notable difference in accuracy, leading us to use the 500 trees as recommended by Belgiu & Drăguţ (2016). Lawrence et al. (2006) note that the use of 500 trees allows the model to stabilize errors before all trees are processed. We test tree depths of 5, 15, 30, 80, and 100. Using tree depths of 5, 80, and 100 resulted in lower model accuracy whereas a maximum depth of 15 and 30 trees results in the highest accuracy; therefore, we choose 30 trees. We also use 1000 samples per class to ensure that a sufficient number of pixels are included in training. Because there is some variation within galactic components, this is particularly important to ensure we are able to train the model on all interclass differences.

To analyze the importance of the parameters, we run the RF algorithm in R Studio (Version 1.1.463; RStudio Team, 2020) using all 38 layers. The GIS program used for MLC, SVM, and RF classifications is not capable of processing a mean decrease in Gini coefficient (MDG) importance plot. Therefore, we make use of the "randomForest" package in R programming language to produce a MDG plot ranking the importance of all layers using the "varImpPlot()" function. The higher the mean decrease in Gini, the more important the parameter is for classification (Koo et al., 2021). MDG also identifies natural subgroups from analysis of all parameters.

### 2.5.4. Classification Groups

From the analysis of the 38 parameters created, we identify the following groups of classifications:

1. HST bands
2. Most important textures
3. Less important textures
4. All eight Haralick textures

5. Distance and HST bands
   6. Band flux ratios
   7. Fraction band ratios
   8. Spectral slope ratios
   9. Top important MDG layers

We perform each classification three times: once for MLC, once for RF, and once for SVM. As a baseline comparison, we classify the original HST BVI imagery. Because there are eight textures in total, we choose to classify the ones identified as most important by the MDG plot. Further, we also classify less important textures to see how the accuracy changes. Along with these, we classify with all eight Haralick textures (Haralick et al., 1973) to determine whether more or less textural information is useful. We also test classification of flux, fraction, and spectral slope band ratios to determine how band ratios contribute to classification of galaxy components. Finally, we classify the most important layers within the top subgroups as identified by the RF MDG plot.

## 2.6. Accuracy Assessment

To assess the accuracy of our model, we split the polygon sites described in section 2.4. into a training set and test set. After testing of 90/10, 80/20, and 70/30 splits, we find that there is no notable difference in accuracy; therefore, we choose to use a 70/30 split representing 70% and 30% of the total number of polygons. The number of polygons and pixels per data set is shown in Table 3. To increase the confidence of our model, we perform classification twice, alternating the sets used for training and testing and averaging the two accuracies.

**Table 3.** Number of polygons and pixels within 70% and 30% of the polygon dataset.

|  | 70% of Dataset | 30% of Dataset |
| --- | --- | --- |
| **Number of Polygons** | 455 | 195 |
| **Number of Pixels** | 170784 | 120605 |

To analyze the prediction power of the MLC, RF, and SVM models, we calculate overall accuracy (OA), user's accuracy (UA), producer's accuracy (PA), and F1 score. OA is defined by the below equation:

$OA$ = correctly classified pixels / total number of pixels in image (12)

Overall accuracy is a commonly used statistic in remote sensing and is useful as a simple measure of the proportion of correctly classified pixels in a map.

The second accuracy statistic UA is described by the equation:

$UA$ = TP / TP + FP (13)

where TP is the true positive and FP is the false positive in the confusion matrix. UA represents the class accuracy for the rows of the confusion matrix. The UA metric is useful for measuring the errors of commission (Congalton & Green, 2019). Similarly, PA defines the class accuracy for the columns of the confusion matrix and is represented by Eq. (14):

$PA$ = TP / TP + FN (14)

The PA metric looks at the errors of omission making it useful for knowing what samples have been omitted from being correctly classified (Congalton & Green, 2019). Both the UA and PA metrics take into account confusion matrix error. Because OA does not account for error, reporting on the UA and PA values is ideal for ensuring the confusion matrix is summarized properly.

The fourth and final accuracy metric, the F1 Score, analyzes both the user's and producer's accuracy statistics of a confusion matrix. The F1 Score for an individual class is calculated as the mean of user's and producer's accuracy by the equation:

$F1$ = 2 * (UA * PA) / (UA + PA) (15)

We also evaluate the F1 Score for the overall map, so we average the individual class F1 Scores (Goutte & Gaussier, 2005).

## 3. Results
### 3.1. Parameter Importance

The Mean Decrease Gini (MDG) plot from RF classification in Figure 8 identifies several groups of importance from the input parameters. Overall, the galaxy center distance is the most important of the 38 total parameters and forms its own group. We also find that Mean texture parameters are the most important textures and form their own subgroup. The third group contains the spiral arm distance, band F814W (infrared) Correlation texture, and the HST band F606W (visual). We perform classification of these three groups (top seven MDG layers). Because both distance parameters and all three Mean texture parameters are within the top three subgroups, we conclude that these are the most useful parameters for galaxy component classification. Angular Second Moment is the least important texture according to the MDG plot. Therefore, we also perform classifications of Mean textures and Angular Second Moment textures as the most and least important textures respectively.

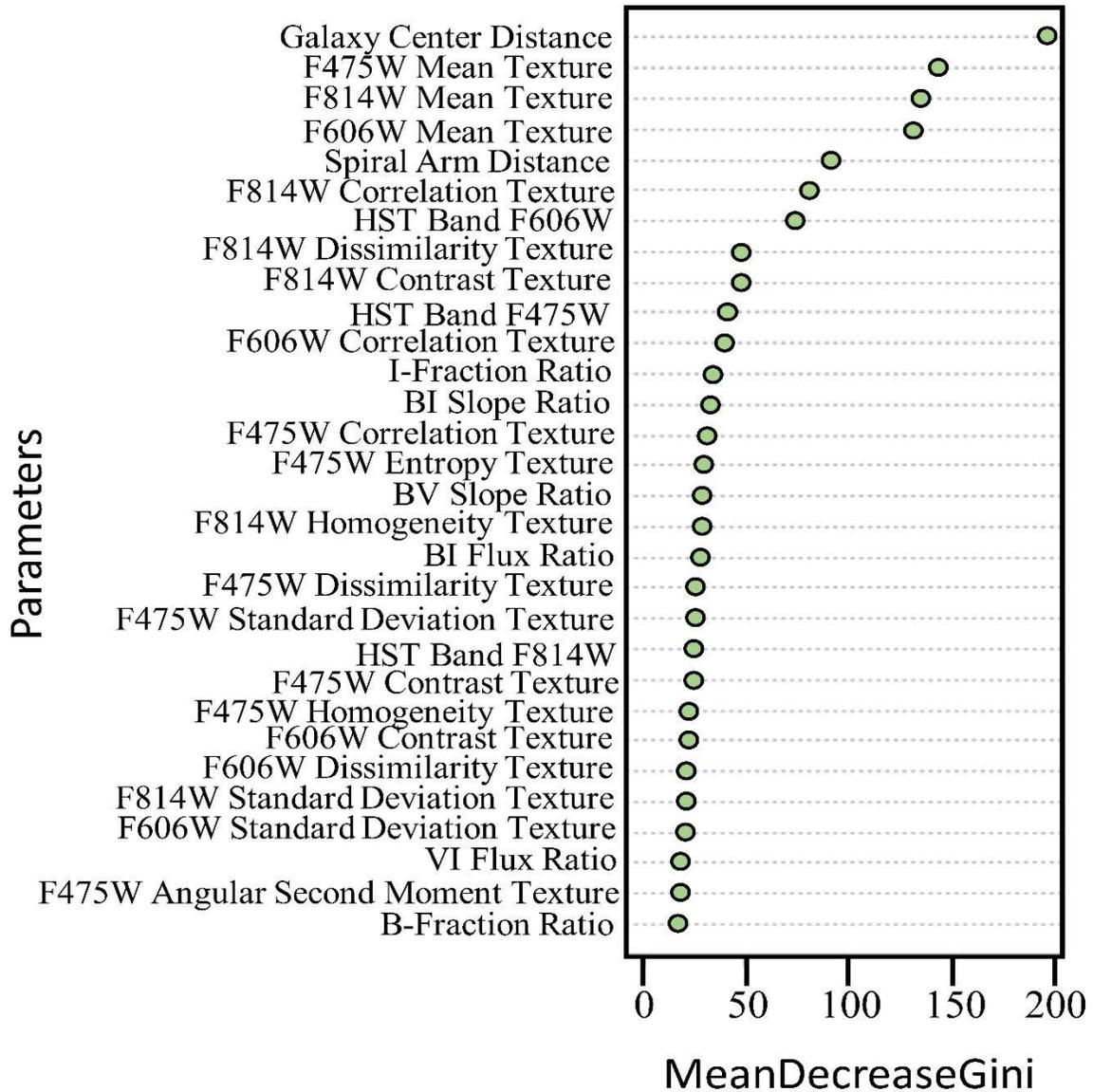

**Figure 8.** MDG plot of the top 30 important input parameters.

Figure 9 shows maps of classifications with the top performing layers from the MDG plot as compared to the classification with HST bands, which we use as a baseline. Because SVM model performs slightly better than the MLC and RF models, we compare SVM classifications when using 70% of the polygon dataset for training. We do not show the classifications using 30% of the polygon dataset for training as there are no noteworthy differences. Classifications with distance layers – HST bands and distance, and top seven MDG layers – results in the highest accuracies, but also exhibits sharp edges throughout the image. On the contrary, classification with mean textures and classification with HST bands exhibit smoother changes in galaxy component membership throughout the map. This contrast is due to the nature of the Euclidean distance rasters where the distance is discrete rather than a continuous surface. However, classification with the top seven layers incorporates both texture and distance along with HST band F606W, reducing effects of the distance layers. The main pitfalls of the use of distance in classification are the lack of pixels classified as old stellar population and the overemphasis of the dust lanes and outer disc. Use of mean texture appears to improve upon classification with HST bands as it does a better job of identifying old stellar populations within the digital imagery. Classification with only the HST bands tends to overemphasize the dust lanes. Similarly, young stellar populations are better defined within the mean texture classification.

    Comparison of all maps reveals that classification of mean texture and classification of the top seven MDG parameters are better at classifying the old stellar populations within the inner disc of the galaxy. However, in the classification of the top seven MDG parameters, the discrete effects of the distance parameters are stronger in the outer disc region as shown by the arrow in the upper inset image (Figure 9). There is a sudden change in galaxy component membership going from young stellar population to dust lanes to outer disc. In the original HST BVI imagery, this is not the case and that there is more diversity of galaxy components within this same region, similar in appearance to the classification with mean textures. The same effect can be seen in the lower inset map where mean texture has the most diversity of galaxy components and therefore best resembles the original HST BVI imagery. However, the classification with HST bands, HST bands and distance, and the top seven MDG parameters does not show the same diversity of galaxy components within the region identified by the arrow in the lower inset map (Figure 9). Therefore, although we find that the classification with top seven MDG parameters results in the highest accuracy, the classification of mean textures is best at classifying the small-scale changes in galaxy component membership throughout the digital imagery.

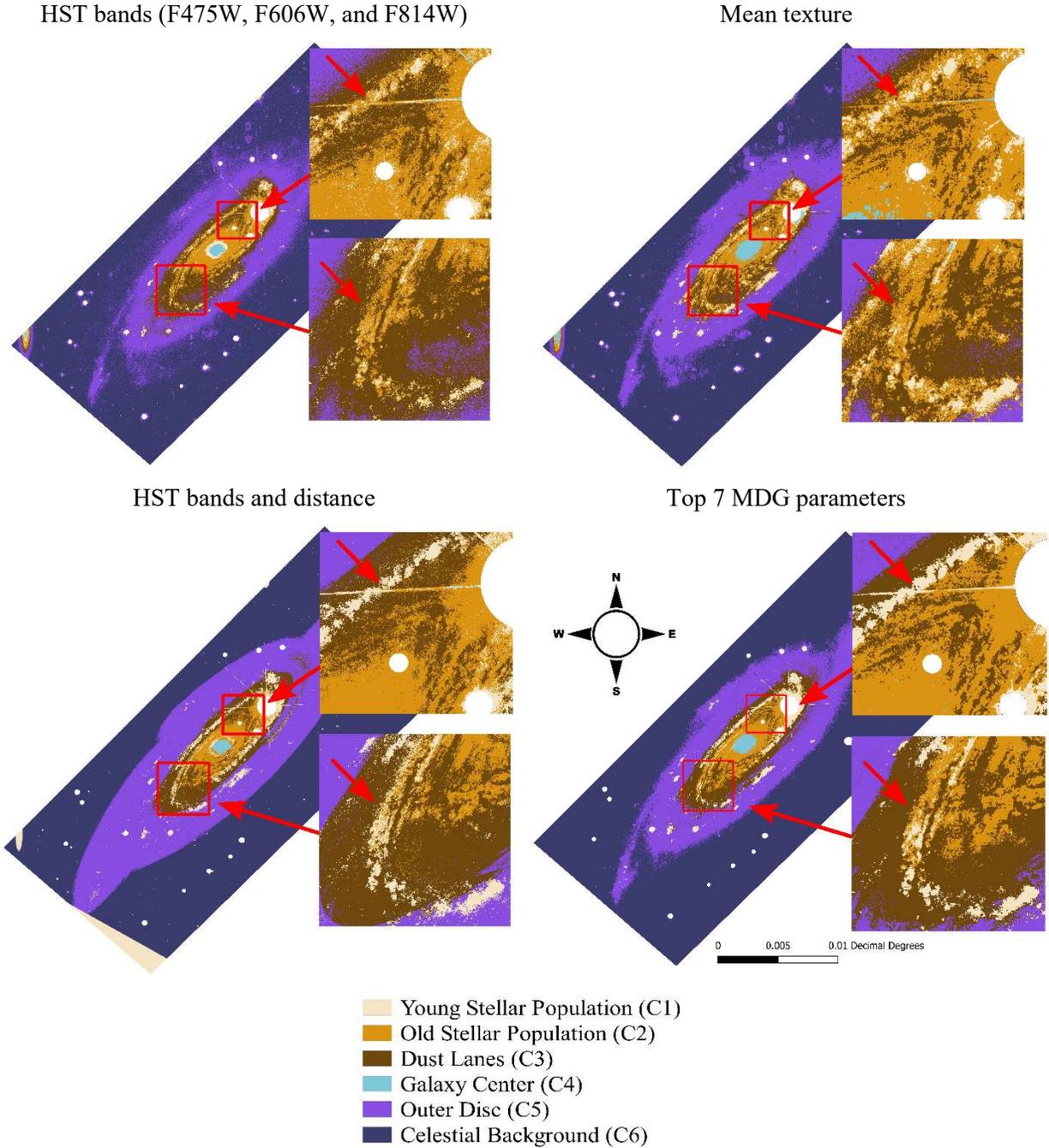

**Figure 9.** SVM classifications using 70% of the polygon set as training and 30% of the polygons as testing. Here east is to the right of the north direction as the imagery is georeferenced in Earth-based coordinates. We include inset maps, represented by red squares, to show the small details within the inner disc of the galaxy.

We include the confusion matrices of the most accurate classification of SVM with the top seven MDG layers in Table 4 and Table 5. A confusion matrix compares the pixels in each class within the test set and the classification results of the pixels within the classified imagery. The diagonal represents the correctly classified pixels for their respective classes and is coloured in green. The column and row coloured in yellow represents the UA and PA respectively. We colour the OA statistic in blue. The row and column titled 'Total' represent the total number of pixels in each class from the original reference data (test set) and the classified image respectively. Misclassifications are the boxes not along the diagonal, excluding the Total, UA, and PA rows and columns as well as the OA. We bold the boxes that represent the most confusion, meaning that these classes have higher rates of misclassification between them in the form of misclassified pixels.

Table 4 shows SVM classification of the top seven layers when using 70% of the polygon set for training while Table 5 shows SVM classification of the top seven layers when using 30% of the polygon set for training. From Table 4, we notice that there is the most confusion between classes C2 (old stellar population) and C3 (dust lanes). This agrees with our expectations as discussed in section 2.4. Similarly, class C3 (dust lanes) and C5 (outer disc) share confusion due to their spectral similarities. Within the inner disc, it is easier to distinguish between the dust lanes and other galaxy components. However, in the outer disc where there are less galaxy components, the HST imagery is darker making it more difficult to visually separate the components within it. Class C1 (young stellar population) and C2 (old stellar population) share a bit of confusion within both tables. However, C1 performs better due to its bluer appearance. The least confusion is present in the C4 (galaxy center) and C6 (celestial background) classes as these are the most spectrally unique parts of the HST imagery.

**Table 4.** Confusion matrix of SVM classification of the top seven layer using 70% of the polygon set for training and 30% for testing. The green shading shows the number of correctly classified pixels for each class, yellow shading shows the PA and UA row and column, and the blue shading shows the OA statistic[2]. Bolded numbers identify the areas of the most confusion between classes.

|  |  | Reference data by expert interpretation |  |  |  |  |  |  |  |
|---|---|---|---|---|---|---|---|---|---|
|  |  | C1 | C2 | C3 | C4 | C5 | C6 | Total | UA |
|  | C1 | 9856 | 55 | 3 | 0 | 9 | 0 | 9923 | 0.99 |
|  | C2 | 185 | 9579 | **696** | 0 | 0 | 0 | 10460 | 0.92 |
| ML | C3 | 46 | **875** | 10270 | 0 | **800** | 0 | 11991 | 0.86 |
| Classification | C4 | 52 | 113 | 0 | 6279 | 0 | 0 | 6444 | 0.97 |
|  | C5 | 0 | 0 | **578** | 0 | 49622 | 0 | 50200 | 0.99 |
|  | C6 | 0 | 0 | 0 | 0 | 9 | 31578 | 31587 | 1.00 |
|  | Total | 10139 | 10622 | 11547 | 6279 | 50440 | 31578 | 120605 |  |
|  | PA | 0.97 | 0.90 | 0.89 | 1.00 | 0.98 | 1.00 |  | 0.96 |

**Table 5.** Confusion matrix of SVM classification of the top seven layer using 30% of the polygon set for training and 70% for testing. The green shading shows the number of correctly classified pixels for each class, yellow shading shows the PA and UA row and column, and the blue shading shows the OA statistic. Bolded numbers identify the areas of the most confusion between classes.

|  |  | Reference data by expert interpretation |  |  |  |  |  |  |  |
|---|---|---|---|---|---|---|---|---|---|
|  |  | C1 | C2 | C3 | C4 | C5 | C6 | Total | UA |
|  | C1 | 27696 | 222 | 46 | 0 | 6 | 0 | 27970 | 0.99 |
|  | C2 | **653** | 16788 | **2073** | 0 | 0 | 0 | 19514 | 0.86 |
| ML | C3 | 90 | **2417** | 27229 | 0 | **4707** | 0 | 34443 | 0.79 |
| Classification | C4 | 0 | 0 | 0 | 9105 | 0 | 0 | 9105 | 1.00 |
|  | C5 | 0 | 0 | 531 | 0 | 36281 | 0 | 36812 | 0.98 |
|  | C6 | 0 | 0 | 0 | 0 | 113 | 42827 | 42940 | 1.00 |
|  | Total | 28439 | 19427 | 29879 | 9105 | 41107 | 42827 | 170784 |  |
|  | PA | 0.97 | 0.86 | 0.91 | 1.00 | 0.88 | 1.00 |  | 0.94 |

## 3.2. Galaxy Component Classification Performance

We present the user's accuracy (UA), producer's accuracy (PA), and F1 Scores in Table 6, Table 7, and Table 8. The ML models are best at predicting galaxy center and celestial background as they exhibit lower rates of confusion than the remaining classes (Figure 7); this is due to the brightness of the galaxy center and the darkness of the celestial background. Class C5, the outer disc, also exhibits high prediction power. According to Table 6, 7, and 8, the old stellar population (C2) and dust lane (C3) classes have the lowest classification accuracy due to the spectral similarities between these classes. For instance, Figure 7 (section 2.4.) demonstrates the confusion between these three classes due to the visual similarities and mixing within the inner disc of the galaxy due to the limited resolution of the HST imagery. The young stellar population (C2) has slightly higher UA and PA values as this class is more spectrally unique than both the old stellar population and dust lanes; by spectrally unique, we refer to the brightness and blueness of the young stars within the galaxy. In contrast, the old stars and dust lanes tend to be redder and dimmer than the young stars.

---
[2] Accuracy statistics are rounded to two decimal places.

Overall, the MLC model best classified the outer disc (C5) and celestial background (C6) classes for each accuracy statistics (Table 6, 7, 8). MLC has low classification accuracy for the old stellar population class (C2). In comparison to the RF and SVM models, MLC class accuracies are comparative to the RF class accuracies more so than to the SVM class accuracies. For instance, in Table 6, MLC and RF have the same average UA of 0.80 for young stellar population (C1) classification. For producer's accuracy in Table 7, MLC model outperforms RF accuracy of classification of dust lanes (C3) and outer disc (C5). The SVM model consistently outperforms individual class accuracies of the MLC model.

**Table 6.** Average of user's accuracy for each class in the MLC, RF, and SVM models.[3] The bolded numbers represent the model with the highest user's accuracy for each class. SVM model outperforms the MLC and RF models.

| Model | User's Accuracy Average | | | | | |
|---|---|---|---|---|---|---|
| | C1 | C2 | C3 | C4 | C5 | C6 |
| MLC | 0.80 | 0.64 | 0.68 | 0.69 | 0.89 | 0.88 |
| RF | 0.80 | 0.70 | 0.65 | **0.87** | 0.91 | 0.93 |
| SVM | **0.84** | **0.77** | **0.73** | 0.83 | **0.92** | **0.94** |

**Table 7.** Average of producer's accuracy for each class in the MLC, RF, and SVM models.[3] The bolded numbers represent the model with the highest producer's accuracy for each class. SVM and RF models outperform the MLC.

| Model | Producer's Accuracy Average | | | | | |
|---|---|---|---|---|---|---|
| | C1 | C2 | C3 | C4 | C5 | C6 |
| MLC | 0.75 | 0.70 | 0.76 | 0.83 | 0.79 | 0.89 |
| RF | **0.82** | 0.71 | 0.70 | **1.00** | 0.77 | **0.96** |
| SVM | 0.80 | **0.72** | **0.78** | **1.00** | **0.81** | **0.96** |

---

[3] We average the accuracies over all nine classifications of each model.

**Table 8.** Average of F1 Score for each class in the MLC, RF, and SVM models.[3] The bolded numbers represent the model with the highest F1 score accuracy for each class. SVM model outperforms the MLC and RF models.

| Model | F1 Score Average | | | | | |
|---|---|---|---|---|---|---|
| | C1 | C2 | C3 | C4 | C5 | C6 |
| MLC | 0.77 | 0.65 | 0.71 | 0.75 | 0.82 | 0.89 |
| RF | 0.79 | 0.68 | 0.67 | **0.93** | 0.81 | **0.95** |
| SVM | **0.81** | **0.72** | **0.73** | 0.90 | **0.85** | **0.95** |

   Figure 10 shows maps of the top four classifications as determined by average of overall accuracy statistics and F1 Scores (Table 9). For the highest accuracy classifications reported in Figure 10 and Table 9, MLC performs with a similar accuracy to the RF and SVM models, but still falls short by about 2% for both OA and F1 score accuracies. From observation of the maps, we note that RF and SVM classifications appear quite similar. The main difference between the resulting maps is that the HST bands and distance classifications create sharper transitions as shown by the arrows in the upper and lower inset maps. The SVM classifications show more small details within the galaxy. Therefore, the RF model tends to emphasize the distance layers the most in both classification types shown in Figure 10. We find that both RF and SVM are useful methods of classification for digital imagery of galaxies.

   Visually, the best models in Figure 10 are those using the top seven MDG parameters. This agrees with the OA and F1 Score as the models using top seven layers result in slightly higher (1%) accuracies than the HST band and distance models. The galaxy center is best identified by the SVM classification with top seven MDG parameters; in the other three classifications in Figure 10, there is some visual confusion between the class membership of the galaxy center and young stellar population. SVM classification with top seven MDG layers is also best at identifying the old stellar population within the inner disc of the galaxy, although confusion with dust lanes remains an issue. Ultimately, the arrows indicated inside the inset maps demonstrate that there is not much difference between the classifications of HST bands and distance and of the top seven MDG parameters. The accuracies of each model, as reported in Table 9, agree with this finding.

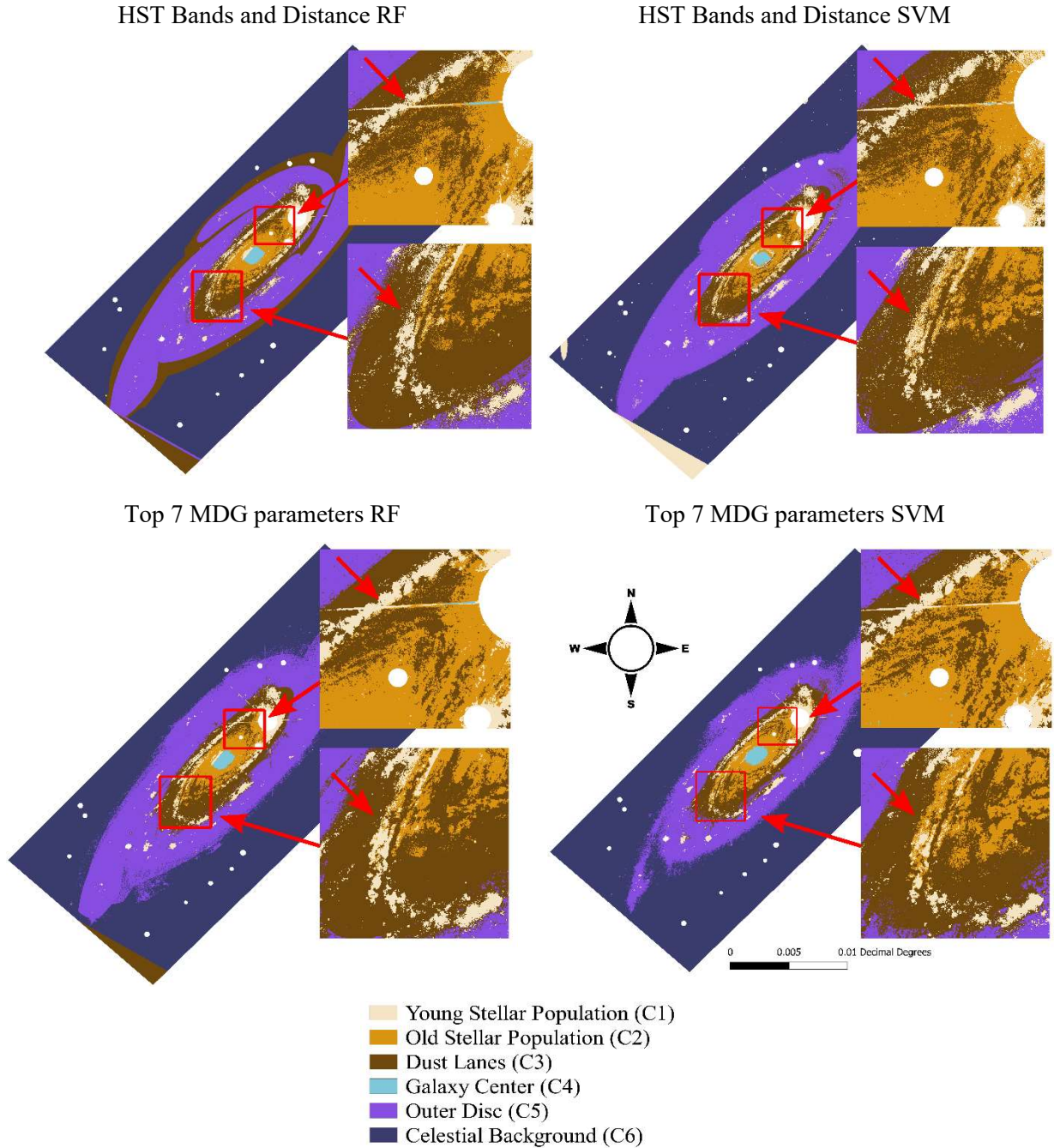

**Figure 10.** Top performing classifications. All maps shown are the product of classifications using 70% of the polygon set as training and 30% as testing. We include inset maps, represented by red squares, to show the fine details within the inner disc of the galaxy.

Table 9. Top performing classifications from Figure 10 and their corresponding accuracy statistics.

|  | HST Bands and Distance RF | HST Bands and Distance SVM | Top 7 MDG parameters RF | Top 7 MDG parameters SVM |
|---|---|---|---|---|
| **OA** | 94.7% | 94.5% | 95.4% | 95.4% |
| **F1 Score** | 93.5% | 93.3% | 94.4% | 94.6% |

### 3.3. Model Performance Summary

We perform nine classifications for each ML model: HST bands, mean texture, angular second moment texture, all textures, distance and HST bands, band flux ratios, fraction band ratios, spectral slope ratios, and top seven MDG layers. Overall, the RF and SVM models result in higher performance than the traditional MLC model with SVM being slightly more successful at predicting galactic components over the nine classifications performed (Table 10). The MLC model performs particularly well with the spectral slope band ratio classification with an average OA of 85.3%. For the same spectral slope classification, RF and SVM models perform with an average OA of 48.1% and 66.7% respectively. Otherwise, RF and SVM consistently outperform the MLC model. Therefore, we suggest that RF and SVM models be used for galaxy component classification.

Table 10. Average accuracies for each model.[4]

| Model | Overall accuracy | F1 Score |
|---|---|---|
| MLC | 80.5% | 80.4% |
| RF | 82.6% | 80.9% |
| **SVM** | **84.9%** | **82.6%** |

### 4. Discussion

In this study, we use MLC, RF, and SVM machine learning models to classify galaxy component membership within HST digital imagery of UGC 2885. Along with HST imagery, we input band ratios derived from HST imagery, textural features derived from HST imagery, and distance layers into the classification to determine the most effective method of class membership prediction. We identify six classes within the digital imagery: young stellar population, old stellar population, dust lanes, galaxy center, outer disc, and celestial background. To analyze accuracy of galaxy component membership, we make use of PA and UA statistics. The classes with the most confusion are young stellar population and old stellar population as well as old stellar population and dust lanes as they share similar spectral appearance. The classes with the least confusion are galaxy center, outer disc, and celestial background. This is expected as these classes are unique in spectral appearance. Among the MLC, RF, and SVM machine learning models used, SVM results in the highest accuracy of galaxy component classification between both PA and UA statistics. SVM also results in the highest accuracy for overall map statistics of OA and F1 Score. The RF models have comparable performance. According to the RF Mean Decrease Gini plot, distance and texture parameters are most useful for galaxy component membership prediction. This finding is confirmed by the high accuracies that classification with distance and textures yields. Finally, a combination of HST bands, texture, and distance results in the highest accuracy. The combined power of several types of information is optimal for galaxy component classification within digital imagery. The success of the SVM and RF models and relatively poor performance of MLC is expected and agrees with results from other recent studies comparing machine learning algorithms in both remote sensing and astronomy (Ghayour et al., 2021; Wang et al., 2021).

---
[4] We average the accuracies over all nine classifications of each model.

      One limitation of our study is the lack of 'ground truth' reference data. Due to this lack of reference data, we use expert visual interpretation to train the models. Since UGC 2885 is 71 Mpc away, some galaxy components are difficult to distinguish. Because of this, we are only able to train the model on pixels whose class membership we are certain. This means that large portions of the galaxy are ignored in accuracy assessment, likely contributing to the high accuracy results. To improve the reliability of accuracy assessment, we recommend classifying galaxies that have some sort of reference data available. Further, galaxies in closer proximity may be more suitable due to the higher resolution of imagery, making it easier to identify galaxy components.

      One manual step in our study is the tracing of spiral arms. Although this method has been used for several purposes in spiral galaxy research (Scheepmaker et al., 2009; Shabani et al., 2018; Bialopetravičius & Narbutis, 2020), it would be infeasible to manually trace spiral arms of many galaxies if using our model to automate rapid galaxy component classification. Research is

being done to automate spiral arm fitting (Davis & Hayes, 2014; Bekki, 2021), so the addition of an automated model for spiral arm fitting into our machine learning model would drastically decrease processing time. Additionally, deprojection of galaxy imagery is unnecessary for spiral arm fitting models due to the low dependence on disc inclination angle (Bekki, 2021). Due to the manual tracing of the spiral arms, there may be some uncertainties present in the final classification of the galaxy components. A more mathematical approach to spiral arm fitting would likely increase accuracy of spiral arm tracing. More research would be needed to determine how spiral arm fitting could affect uncertainty of the final classification.

For future study of galactic components, we find that use of textures may improve classification accuracy. We recommend that texture analysis be experimented with further to explore its full potential for astronomical research. We recommend that texture also be tested on imagery of other deep space celestial phenomena such as irregular galaxies that exhibit tonal variation and patterns within digital imagery. One limitation of texture is that it is not as effective when applied to low resolution imagery. We recommend texture be used for high resolution HST imagery or for nearby celestial phenomena. For instance, textural features would be useful for upcoming telescopes such as Euclid, Roman Space Telescope, and James Webb Space Telescope that are expected to produce high-resolution imagery comparable to that of HST (Gardner et al., 2006; Laureijs et al., 2011; Spergel et al., 2015). Along with texture, we also recommend the use of distance measures for classification of galaxy components, in conjunction with other parameters such as texture and HST bands to achieve the optimal accuracy.

The machine learning models used present their own set of limitations. One downfall of the SVM algorithm is that it ignores training samples that do not support the hyperplane (Foody & Mathur, 2006). This might cause classes with a wide range of pixel values to be classified poorly. Presence of extraneous features may negatively affect the performance of the model. We recommend that the features be analyzed to identify any outlier features to avoid a significant decrease in accuracy of the model (Baron, 2019). MLC is not able to handle data with a non-normal distribution as it attempts to define a unique probability density function for each class (Otukei & Blaschke, 2010). To combat this, data with normal distribution should be used for MLC classification. The HST data used here has normally distributed flux so is compatible with the MLC algorithm. A disadvantage to the RF algorithm is that it cannot handle datasets with imbalanced training samples (Dalponte et al., 2013). Astronomical data has noise present and the ML methods used here fail to account for that. ML algorithms that account for uncertainties of both features and labels have only recently been developed (e.g., Reis et al., 2019). To take astronomical noise into account, probability distribution functions are created for the features and labels. This improves accuracy of RF classification by 10-30% (Reis et al., 2019). Currently, RF models in widely-used packages and GIS programs are not capable of accounting for uncertainty in input data. Incorporating ML algorithms that account for uncertainty would be a natural extension of our model and would improve accuracy of classification.

Our ML model is successful at classifying galaxy components within a nearby spiral galaxy, UGC 2885. Dissecting these fine structural details within galaxies is important for understanding formation and evolution of galaxies (Lingard et al., 2020; Peng et al., 2002). UGC 2885 exhibits an extended disc with a sparse population of young stars that we classify using visible HST imagery. We also notice that a large portion of the stellar matter we can see is located in the inner disc. Further research on galaxies of different spiral forms and life cycle stages is needed to fully understand the secular spatial and temporal changes of galaxy component distribution. Classification using ours or similar models helps to automate that process.

## 5. Conclusions

In this study, we proposed a machine learning (ML) approach for galaxy mapping of UGC 2885 using high-resolution Hubble Space Telescope (HST) digital imagery. We compare three ML models: maximum likelihood classifier (MLC), random forest (RF), and Support Vector Machine (SVM). ML is successful at mapping galaxy components: RF and SVM models are found to have the strongest classification power whereas MLC performance is slightly inferior. The ML models successfully classify all identified components within the digital imagery, with the most confusion shared between the dust lanes and old stellar populations within the galaxy. The young stellar population, galaxy center, outer disc, and celestial background are the best classified by the ML models and therefore have the least confusion. From analysis of parameter importance, distance and mean textural parameters are the most important for galaxy component classification. The best performing models were those using the top seven mean decrease Gini parameters, a combination of distance, textural features derived from HST imagery, and HST digital imagery data, making this method particularly important. Further research could determine the full potential of textural analysis for study of galaxies and other celestial phenomena. These findings are relevant for the soon to be launched Euclid, Roman Space Telescope, and James Webb Space Telescope as these telescopes will provide an abundance of high-resolution data similar to the HST data used in this study. Our research demonstrates that the automation of mapping the fine galaxy component structures within digital imagery is feasible.

## Acknowledgements

We would like to thank Dr. Philip Stooke for providing constructive feedback and advice throughout the research process.

This work is made possible by funding from NSERC Discovery Grant awarded to Dr. Wang and to Dr. Barmby, respectively.

This work has made use of data from the European Space Agency (ESA) mission Gaia (https://www.cosmos.esa.int/gaia), processed by the Gaia Data Processing and Analysis Consortium (DPAC, https://www.cosmos.esa.int/web/gaia/dpac/consortium). Funding for the DPAC has been provided by national institutions, in particular the institutions participating in the Gaia Multilateral Agreement.

Pedersen, K.S., Stensbo-Smidt, K., Zirm, A., et al., 2013. Shape Index Descriptors Applied to Texture-Based Galaxy Analysis, in: 2013 IEEE International Conference on Computer Vision. Presented at the 2013 IEEE International Conference on Computer Vision (ICCV), IEEE, Sydney, Australia, pp. 2440–2447. https://doi.org/10.1109/ICCV.2013.303

Peng, C.Y., Ho, L.C., Impey, C.D., et al., 2002. Detailed Structural Decomposition of Galaxy Images. The Astronomical Journal 124, 266–293. https://doi.org/10.1086/340952

QGIS 3.16. (2021). Geographic Information System Developers Manual. <docs.qgis.org/3.16/en/docs/developers_guide/index.html>

Reis, I., Baron, D., Shahaf, S., 2019. Probabilistic Random Forest: A Machine Learning Algorithm for Noisy Data Sets. AJ 157, 16. https://doi.org/10.3847/1538-3881/aaf101

Richards, J.A., Jia, X., 1999. Remote Sensing Digital Image Analysis. Springer Berlin Heidelberg, Berlin, Heidelberg. https://doi.org/10.1007/978-3-662-03978-6

RStudio Team, 2020. RStudio: Integrated Development for R; RStudio: Boston, MA, USA. www.rstudio.com/

Scheepmaker, R.A., Lamers, H.J.G.L.M., Anders, P., et al., 2009. The spatial distribution of star and cluster formation in M 51. A&A 494, 81–93. https://doi.org/10.1051/0004-6361:200811068

Schutter, A., Shamir, L., 2015. Galaxy morphology — An unsupervised machine learning approach. Astronomy and Computing 12, 60–66. https://doi.org/10.1016/j.ascom.2015.05.002

Seigar, M.S., James, P.A., 1998. The structure of spiral galaxies — II. Near-infrared properties of spiral arms. Monthly Notices of the Royal Astronomical Society 299, 685–698. https://doi.org/10.1046/j.1365-8711.1998.01779.x

Shabani, F., Grebel, E.K., Pasquali, A., et al., 2018. Search for star cluster age gradients across spiral arms of three LEGUS disc galaxies. Monthly Notices of the Royal Astronomical Society 478, 3590–3604. https://doi.org/10.1093/mnras/sty1277

Shamir, L., 2009. Automatic morphological classification of galaxy images. Monthly Notices of the Royal Astronomical Society 399, 1367–1372. https://doi.org/10.1111/j.1365-2966.2009.15366.x

Shamir, L., 2021. Automatic identification of outliers in Hubble Space Telescope galaxy images.
41

# Figure Captions

**Figure 1.** HST colour composite map of UGC 2885 where the F475W band is displayed in blue, the F606W band is displayed in green, and the F814W band is displayed in red. This image is in celestial orientation; therefore, east is towards the left rather than to the right of the north direction. Parsecs refers to a measure of distance equalling approximately 3.26 light years. Image credit: NASA, ESA, and B.W. Holwerda (University of Louisville).

**Figure 2.** UGC 2885 blue-green band imagery (F475W): a) original HST imagery showing the projected galaxy with an inclination of 74°; b) deprojected image of UGC 2885 – stretched vertically by 363%.

**Figure 3.** Flowchart showing methods used in this study.

**Figure 4.** Band F606W Haralick textures with 20 brightest foreground stars masked.

**Figure 5.** Spiral arms and galaxy center shown in red and blue respectively. The background image shows the deprojected UGC 2885 F475W blue-green band (Holwerda et al., 2020). Both distance rasters are reprojected to the galaxy's 72° inclination.

**Figure 6.** Examples of training site selection using the classification scheme with six classes. Background image is a RGB colour composite of HST band data of UGC 2885: F475W band is displayed in blue, the F606W band is displayed in green, and the F814W band is displayed in red.

**Figure 7.** Scatterplot of classification scheme pixels and their respective digital number (DN) values within the visual and near-infrared HST bands.

**Figure 8.** MDG plot of the top 30 important input parameters.

**Figure 9.** SVM classifications using 70% of the polygon set as training and 30% of the polygons as testing. Here east is to the right of the north direction as the imagery is georeferenced in Earth-based coordinates. We include inset maps, represented by red squares, to show the small details within the inner disc of the galaxy.

**Figure 10.** Top performing classifications. All maps shown are the product of classifications using 70% of the polygon set as training and 30% as testing. We include inset maps, represented by red squares, to show the fine details within the inner disc of the galaxy.